\documentclass[aip,preprint]{revtex4-2}
\usepackage{tikz}
\usepackage{pgfplots}
\usepackage{pgfplotstable}
\pgfplotsset{width=10cm,compat=1.9}
\usepackage[cp1251]{inputenc}
\usepackage[T2A]{fontenc}
\usepackage[english]{babel}
\usepackage{amssymb,latexsym,amsmath,amscd}
\usepackage{graphicx,color,framed}
\usepackage{bm}
\usepackage{appendix}
\usepackage{tabu}
\usepackage{tikz}
\usepackage{xcolor}
\usepackage{siunitx}
\usepackage{empheq}
\usepackage{float}
\usepackage{subcaption}
\usepackage{xr}
\usepackage{siunitx}
\usepackage{microtype}
\usepackage{amsfonts}
\usepackage{gensymb}
\usepackage[normalem]{ulem}
\graphicspath{{figures/}}

\makeatletter
\newcommand*{\addFileDependency}[1]{
  \typeout{(#1)}
  \@addtofilelist{#1}
  \IfFileExists{#1}{}{\typeout{No file #1.}}
}
\makeatother

\begin{document}

\author{\firstname{Yury A.} \surname{Budkov}}
\email[]{ybudkov@hse.ru}
\affiliation{Laboratory of Multiscale Modeling of Molecular Systems, G.A. Krestov Institute of Solution Chemistry of the Russian Academy of Sciences, 153045, Akademicheskaya st. 1, Ivanovo, Russia}
\affiliation{Laboratory of Computational Physics, HSE University, Tallinskaya st. 34, 123458 Moscow, Russia}
\affiliation{School of Applied Mathematics, HSE University, Tallinskaya st. 34, 123458 Moscow, Russia}
\author{\firstname{Nikolai N.} \surname{Kalikin}}
\affiliation{Laboratory of Multiscale Modeling of Molecular Systems, G.A. Krestov Institute of Solution Chemistry of the Russian Academy of Sciences, 153045, Akademicheskaya st. 1, Ivanovo, Russia}
\author{\firstname{Petr E.} \surname{Brandyshev}}
\affiliation{Laboratory of Computational Physics, HSE University, Tallinskaya st. 34, 123458 Moscow, Russia}
\affiliation{School of Applied Mathematics, HSE University, Tallinskaya st. 34, 123458 Moscow, Russia}
\title{Thermomechanical Approach to Calculating Mechanical Stresses in Inhomogeneous Fluids and Its Applications to Ionic Fluids}
\begin{abstract}
This extended article presents a thermomechanical approach for calculating the stress tensor from the thermodynamic potential of inhomogeneous fluids and some applications to ionic fluids. The technique, based on the invariance of the fluid's thermodynamic potential with respect to spatial transformations of translation and rotation, offers an alternative to the general covariant approach developed by two of the authors. We apply this technique to both pure mean-field theories of fluids in general and a theory that includes thermal fluctuations of the order parameter, using the example of ionic fluids. Additionally, we apply the thermomechanical approach to fluid models with vector order parameters, such as liquid dielectrics. For this case, we obtain a general expression for the stress tensor. Furthermore, we discuss specific issues related to the calculation of disjoining pressure in ionic fluids confined in nanoscale slit-like pores with metal or dielectric walls, using the Coulomb gas model. To test the robustness of the proposed approach, we reproduce a number of known results from the statistical theory of inhomogeneous fluids and obtain several new ones.
\end{abstract}

\maketitle

\section{Introduction}

Understanding the mechanical behavior of liquid-phase electrolytes confined within solid nanostructures is important for various applications, such as batteries, supercapacitors, and electrowetting devices. In order to study these systems effectively, it is necessary to calculate the mechanical stress using a stress tensor alongside the local electrostatic potential and ionic concentration distributions. By determining the local stress tensor based on specific mean field equations one can estimate such properties, as solvation pressure and shear stresses. They play a significant role in predicting the deformation of porous materials, which is essential for optimizing the performance of energy storage devices using microporous electrodes saturated with liquid-phase electrolytes~\cite{koczwara2017situ,gor2024drives,kolesnikov2022electrosorption,kondrat2023theory,nesterova2024role}. Additionally, the stress tensor is essential for determining the macroscopic force acting on charged macroscopic conductors or dielectrics immersed in liquid-phase electrolytes. Thus, a first-principles approach for calculating the stress tensor in inhomogeneous ionic liquids is relevant to practical applications. Specifically, being able to calculate the components of the stress tensor allows us to understand solvation and shear stresses in ionic liquid films of arbitrary geometry~\cite{kolesnikov2021models}. Understanding the difference between normal and tangential stress at the ionic liquid-electrode interface allows us to calculate the surface energy, which in turn is important for modeling electrowetting phenomena~\cite{marinescu2010electrowetting,monroe2007distinctive}.

One common technique used to analyze microscopic stresses is the Irving-Kirkwood method~\cite{shi2023perspective,rusanov2001condition,rusanov2001three}. Averaging the Irving-Kirkwood microscopic stress tensor over different microstates of a system provides valuable insights, though this process can be challenging~\cite{shi2023perspective}. This method can be effectively implemented within the framework of molecular dynamics and Monte Carlo simulations~\cite{shi2023perspective,cormier2001stress,brown1995general}. However, it is less convenient for modeling ionic fluids within the framework of mean field theory-based continuous models.

A recent alternative to the Irving-Kirkwood technique for the case of continuous models of ionic fluids has been proposed in a series of papers~\cite{budkov2022modified,brandyshev2023noether,budkov2023variational,brandyshev2023statistical}. This new methodology involves the derivation of the stress tensor from the grand thermodynamic potential. By treating the grand thermodynamic potential as a functional of appropriate order parameters and applying Noether's theorems~\cite{noether1971invariant,hermann2021noether,hermann2022force,hermann2022noether,sammuller2023noether} to this functional, this approach, known as the {\sl thermomechanical approach} (TMA)~\cite{budkov2024surface}, integrates concepts from both thermodynamics and continuous media mechanics, combining the grand thermodynamic potential with the stress tensor.

Let us briefly discuss the insights gained from the TMA for different ionic fluids. In the paper~\cite{budkov2022modified}, Noether's first theorem was utilized to analyze the grand thermodynamic potential of an ionic fluid as a functional of the electrostatic potential. This established the local mechanical equilibrium condition by defining it in terms of a symmetric stress tensor, which includes both the Maxwell electrostatic stress tensor and the hydrostatic isotropic stress tensor. 
This equilibrium condition was then extended to include scenarios with external potential forces acting on ions, leading to the derivation of a comprehensive analytical expression for the electrostatic disjoining pressure of an ionic fluid confined within a charged nanopore slit. This extended the well-known DLVO expression to encompass various reference models of ionic fluids. Furthermore, the authors of paper~\cite{brandyshev2023noether} proposed \textit{a general covariant approach} based on Noether's second theorem, which allowed them to derive the symmetric stress tensor from a grand thermodynamic potential for any model of inhomogeneous fluid. This approach is reminiscent of the one used by David Hilbert in the general theory of relativity to derive the energy-momentum tensor from the action functional~\cite{earman1978einstein,weinberg1972gravitation,landau2013classical}. The study applied this method to various models of inhomogeneous ionic fluids that incorporate electrostatic or short-range correlations related to packing effects. This led to the derivation of analytical expressions for the symmetric stress tensors of Cahn-Hilliard-like~\cite{blossey2017structural,vasileva2023theory}, Bazant-Storey-Kornyshev~\cite{bazant2011double}, and Maggs-Podgornik-Blossey ~\cite{blossey2017structural} models.
In a related work~\cite{budkov2023variational}, the authors extended previously developed mean-field formalism within the variational field theory of ionic fluids~\cite{wang2010fluctuation_,lue2006variational_} to incorporate electrostatic correlations of ions. Utilizing the general covariant approach from ref.~\cite{brandyshev2023noether}, they derived a total stress tensor that incorporated these electrostatic correlations by introducing a term depending on the autocorrelation tensor function of local electric field fluctuations near the mean-field configuration. 
By using this total stress tensor and applying the mechanical equilibrium condition, the authors derived a general expression for the disjoining pressure of ionic liquids confined in a slit-like pore. In terms of applications, the TMA has been implemented to model the behavior of solvation pressure in nanopores containing liquid-phase electrolytes~\cite{kolesnikov2022electrosorption,vasileva2023theory} and polyelectrolytes~\cite{budkov2023macroscopic,budkov2023dielectric}, as well as to describe the surface tension of electrolyte solutions at the interfaces with air and dodecane~\cite{budkov2024surface}. Notably, a recent paper~\cite{ruixuan2023electrostatic} utilized the total stress tensor to calculate the force of interaction between charge-regulated colloidal particles immersed in an electrolyte solution. 
It is also worth noting the recent paper~\cite{chen2022porous} which employs a hybrid quantum-classical model to investigate the separation-dependent disjoining pressure between two metal surfaces in an electrolyte solution under surface potential control. The authors proposed an approach to calculating the disjoining pressure based on the stress tensor using the macroscopic electrohydrodynamic method, ultimately leading to the equilibrium state. Based on the developed approach, the authors find that the pressure between the surfaces transitions from a long-range electrostatic interaction, which can be either attractive or repulsive, depending on the charging conditions of the surfaces, to a stronger short-range van der Waals attraction, and then even stronger Pauli repulsion, due to the redistribution of electrons.

It is also instructive to draw reader's attention to related studies that focus on deriving and explicitly evaluating electrostatic normal stress in Coulomb fluids confined in slit-like pores, exploring both weak-coupling and strong-coupling limits~\cite{buyukdagli2023impact,jho2008strong,dean2003field,moreira2001binding}. It is important to note that several papers ~\cite{kendon2001inertial,kruger2018stresses} have proposed approaches based on Ginzburg-Landau theory for calculating local mechanical stresses in simple liquids and colloidal solutions, both in equilibrium and beyond.

The consistent TMA~\cite{brandyshev2023noether,budkov2023variational,brandyshev2023statistical}, as described, is based on a general covariant approach that is rather complex and requires the use of mathematical tools such as Riemannian geometry and tensor analysis. This limits the applicability of the TMA for many physical chemists and chemical engineers who may not be familiar with these techniques. Therefore, having an alternative method that does not rely on differential geometry would be beneficial. In our recent work~\cite{brandyshev2023statistical}, we proposed an approach based on symmetry considerations, utilizing statistical field theory within the functional Legendre transformation formalism. However, these ideas have not yet been systematically presented. This paper aims to fill this gap by presenting the details of our technique, alternative to the sophisticated general covariant approach. We will focus on two practically important cases: (1) the mean field theory of ionic fluids, described by the thermodynamic potential density with spatial derivatives of the order parameter up to the second order, and (2) the theory of the Coulomb gas, which incorporates thermal fluctuations of the electrostatic potential within the framework of variational field theory. As we will see below, the first case can be considered as a part of a broader discussion of the mean-field approach to stress tensors in inhomogeneous fluids, which are described by a scalar or vector order parameter. For both cases, we will present a method for calculating the symmetric stress tensor from the thermodynamic potential or partition function, based on its spatial symmetries, such as translations and rotations.

\section{Stress tensor in inhomogeneous fluids. Mean field theory}

In this section, we will derive the stress tensors for inhomogeneous fluids described by a mean-field functional of the Grand thermodynamic potential (GTP). The potential is defined on scalar and vector order parameters and contains their spatial derivatives up to the second order. We will apply this general approach to the ionic fluids as an example.

\subsection{Stress tensor for thermodynamic potentials, containing first derivatives of the order parameters}

\subsubsection{Stress tensor}
Let us discuss a method that allows us to derive the stress tensor of an inhomogeneous fluid system from the thermodynamic potential functional within the mean-field approximation -- {\sl a thermomechanical approach} (TMA). We consider a general case of an inhomogeneous fluids whose GTP can be represented in the framework of the mean-field approximation as the following functional
\begin{equation}
\Omega=\int d\bold{r} \,\omega(\psi(\bold{r}),\partial \psi(\bold{r})),
\end{equation}
where the GTP density depends on the coordinates through the scalar order parameter $\psi$ and its first coordinate derivatives $\partial_{i}\psi=\partial\psi/\partial x_{i}$. The physical situations describing such kind of GTP densities can be found in Refs.~\cite{kruger2018stresses,brilliantov2020molecular,maggs2016general,budkov2022modified}. It is important to note example of ionic fluid for which the GTP density has the following form \cite{budkov2022modified,maggs2016general}
\begin{equation}
\label{IF_MF}
\omega=-\frac{\varepsilon \left(\nabla \psi\right)^2}{8\pi}-P(\{\bar{\mu}_{\alpha}\}),~\alpha=1,2,\dots,s,
\end{equation}
where $\varepsilon$ is the permittivity of medium, $\alpha$ represents the different types of ionic species, and $\bar{\mu}_\alpha(\mathbf{r}) = \mu_\alpha - q_\alpha \psi(\mathbf{r})$ is the intrinsic chemical potential for the $\alpha$th ion with charge $q_{\alpha}$; the order parameter, $\psi(\mathbf{r})$, in this case represents the local electrostatic potential and $P(\{ \bar{\mu}_\alpha\})$ is the local osmotic pressure of ions.

Now, let us subject the system to an infinitesimal constant translation by the vector $\bold{h}$:
\begin{equation}
\bold{r}^{\prime}=\bold{r}+\bold{h}.
\end{equation}
The first-order variation of the thermodynamic potential with respect to $\bold{h}$ has the following form
\begin{equation}
\delta\Omega=\int d\bold{r} \left(h_{i}\frac{\partial\omega}{\partial x_i}+\delta\omega\right).
\end{equation}
The first term in the integrand describes the variation of $\omega$ with respect to changes in coordinates, while the second term,
\begin{equation}
\label{1_}
\delta\omega=\frac{\partial \omega}{\partial \psi}\delta \psi +\frac{\partial \omega}{\partial(\partial_{i}\psi)}\delta\partial_{i}\psi,
\end{equation}
describes the variation of $\omega$ due to the change of the function $\psi(\bold{r})$. The partial coordinate derivative in the first term implies
\begin{equation}
\label{2_}
\frac{\partial\omega}{\partial x_i}=\left(\frac{\partial\omega}{\partial x_i}\right)_{\psi}+\frac{\partial\omega}{\partial\psi}\partial_{i}\psi+\frac{\partial\omega}{\partial\partial_{k}\psi}\partial_{i}\partial_k\psi,
\end{equation}
where the partial derivative $({\partial \omega}/{\partial x_i})_{\psi}$ is calculated as a derivative of an explicit function at $\psi = const$. In considered case this derivative is zero. Note that here and further, summation by repeating spatial indices is assumed. Then, taking into account the equality
\begin{equation}
\label{3_}
\delta\psi(\bold{r}) =\psi^{\prime}(\bold{r})-\psi(\bold{r})=-h_{k}\partial_{k}\psi(\bold{r}),
\end{equation}
following from the invariance of the order parameter under the change of the coordinate grid, $\psi^{\prime}(\bold{r}^{\prime})=\psi(\bold{r})$, and the identity
\begin{equation}
\frac{\partial \omega}{\partial(\partial_{i}\psi)}\partial_{i}(\delta\psi)=\partial_{i}\left(\delta\psi \frac{\partial \omega}{\partial(\partial_{i}\psi)}\right)- \delta\psi\partial_{i}\left(\frac{\partial \omega}{\partial(\partial_{i}\psi)}\right),
\end{equation}
we obtain
\begin{equation}
\label{deltaOmega}
\delta\Omega = h_{k}\int d\bold{r}\left(\partial_{i}\sigma_{ik}-\partial_{k}\psi\frac{\delta\Omega}{\delta\psi}\right),
\end{equation}
where we have introduced the stress tensor
\begin{equation}
\label{sigma_tens}
\sigma_{ik}=\omega\delta_{ik}-\frac{\partial \omega}{\partial(\partial_{i}\psi)}\partial_k\psi,
\end{equation}
which was first introduced in~\cite{kruger2018stresses}, in the context of the theory of simple fluids, and the variational derivative
\begin{equation}
\frac{\delta\Omega}{\delta\psi}=\frac{\partial \omega}{\partial \psi}-\partial_{l}\frac{\partial \omega}{\partial(\partial_{l}\psi)}.
\end{equation}

Due to the fact that the GTP should be invariant under translation of the system as a whole by a constant vector, i.e., $\delta\Omega=0$, and given that equality (\ref{deltaOmega}) holds for any arbitrary fluid volume, we arrive at the following relation:
\begin{equation}
\partial_{i}\sigma_{ik}=\partial_{k}\psi\frac{\delta\Omega}{\delta\psi}.
\end{equation}
Thus, in the equilibrium configuration of the system, i.e. when the order parameter satisfies corresponding Euler-Lagrange equation, ${\delta\Omega}/{\delta\psi}=0$, the condition of mechanical equilibrium of the fluid is fulfilled
\begin{equation}
\label{partial_sigma}
\partial_{i}\sigma_{ik}=0.
\end{equation}

Note that for the GTP density (\ref{IF_MF}) of ionic fluid the total stress tensor takes the form~\cite{budkov2022modified}
\begin{equation}
\sigma_{ik}=\sigma_{ik}^{(h)}+\sigma_{ik}^{(M)},
\end{equation}
where $\sigma_{ik}^{(h)}=-P\delta_{ik}$ is the hydrostatic stress tensor and
\begin{equation}
\label{sigma_ik_2}
\sigma_{ik}^{(M)}=\frac{\varepsilon}{4\pi}\left(\mathcal{E}_{i}\mathcal{E}_{k}-\frac{1}{2}\mathcal{E}^2\delta_{ik}\right)
\end{equation}
denotes the Maxwell stress tensor with the local electric field components $\mathcal{E}_i=-\partial_{i}\psi$.

It is also important to note that if the GTP density depends on $m$ scalar \textit{order parameters} $\psi^{(s)}$ (see, for instance, \cite{budkov2023macroscopic,vasileva2023theory,blossey2017structural}), $s=1,2,...,m$, and their first partial derivatives, then the expression for the stress tensor have the analogous form:
\begin{equation}
\label{sigma_tens__}
\sigma_{ik}=\omega\delta_{ik}-\sum\limits_{s}\frac{\partial \omega}{\partial(\partial_{i}\psi^{(s)})}\partial_k\psi^{(s)}.
\end{equation}

If the GTP density depends explicitly on the coordinates through the potential of external field $w(\bold{r})$ acting on the fluid particles, with using eqs. 
(\ref{1_}-\ref{3_}) the relation (\ref{deltaOmega}) can be written as
\begin{equation}
\label{deltaOmega_2}
\delta\Omega = h_{k}\int d\bold{r}\left(\partial_{i}\sigma_{ik}-\partial_{k}\psi\frac{\delta\Omega}{\delta\psi}\right)=h_{k}\int d\bold{r}\left(\frac{\partial\omega}{\partial x_k}\right)_{\psi},
\end{equation}
which can be rewritten as follows
\begin{equation}
h_k\int d\bold{r}\left(\partial_{i}\sigma_{ik}-\partial_{k}\psi\frac{\delta\Omega}{\delta\psi}-n\partial_{k}w\right)=0,
\end{equation}
where we took into account that the GTP density $\omega$ depends on coordinates only via $w(\bold{r})$, which is the case with the mean-field approximation~\cite{blossey2017structural,maggs2016general,vasileva2023theory,budkov2022modified,budkov2023macroscopic}, so that
\begin{equation}
\left(\frac{\partial\omega}{\partial x_k}\right)_{\psi}=\frac{\partial \omega}{\partial w}\frac{\partial w}{\partial x_k}=n \frac{\partial w}{\partial x_k}
\end{equation}
with the average number density of the fluid, $n(\bold{r})={\partial \omega}/{\partial w(\bold{r})}$. In particular, in the case of ionic fluids with the GTP density (\ref{IF_MF}) explicit dependence on coordinate is provided the intrinsic chemical potentials of ions $\bar{\mu}_{\alpha}=\mu_{\alpha}-q_{\alpha}\psi - w_{\alpha}$, where $w_{\alpha}$ is the external potential acting on $\alpha$th ion. Thus, due to the arbitrariness of translation vector $\bold{h}$, we can conclude that
\begin{equation}
\label{partial_sigma}
\partial_{i}\sigma_{ik}=\partial_{k}\psi\frac{\delta\Omega}{\delta\psi}+n\partial_{k}w.
\end{equation}

Taking into account that $\delta\Omega/\delta \psi=0$, we obtain the condition of mechanical equilibrium of the fluid in the presence of volume forces
\begin{equation}
\partial_{i}\sigma_{ik}-n\partial_{k}w=0.
\end{equation}

\subsubsection{Proof that tensor $\sigma_{ik}$ is stress tensor}
Now, we need to prove that the tensor $\sigma_{ik}$ for the considered types of GTPs is indeed a stress tensor. Let us subject the system to an inhomogeneous infinitesimal shift according to the rule
\begin{equation}
\bold{r}^{\prime}=\bold{r}+\bold{u}(\bold{r}),
\end{equation}
where $\bold{u}$ is the vector field of displacement. The only difference between this and the previous evaluations is that, instead of a translation, the deformation is changing an infinitesimal volume according to a specific rule~\cite{landau2012theory}:
\begin{equation}
d\bold{r}^{\prime}=\left(1+\partial_{i}u_{i}\right)d\bold{r}.
\end{equation}
Considering that the local electrostatic potential satisfies the Euler-Lagrange equation, ${\delta\Omega}/{\delta\psi}=0$, we obtain the variation of the GTP in the linear approximation over displacement $\bold{u}$ as follows
\begin{equation}
\delta\Omega =\int d\bold{r}\, \partial_{i}\left(u_{i}\sigma_{ik}\right)=\int d\bold{r}\,u_{ik}\sigma_{ik}
\end{equation}
which leads us to the well-known thermodynamic definition of the stress tensor~\cite{landau2012theory}
\begin{equation}
\label{sigma_therm}
\sigma_{ik}=\frac{\delta\Omega}{\delta u_{ik}}\bigg{|}_{u_{ik}=0},
\end{equation}
\begin{equation}
u_{ik}=\frac{1}{2}\left(\partial_{i}u_{k}+\partial_{k}u_{i}\right).
\end{equation}
As mentioned, we assumed the symmetry of the stress tensor $\sigma_{ik}$ during the proof. However, this assumption does not guarantee that the resulting tensor (\ref{sigma_tens__}) will be symmetric. In the case of fluids described by GTPs containing only first derivatives of the order parameter, we will show below that the stress tensor is automatically symmetric.

\subsection{Disjoining pressure}
It is now constructive to discuss how we can obtain the expression for the disjoining pressure in the fluid confined in the slit charged pore, using the expression derived in the previous section for the total stress tensor. Despite the fact that we have already discussed this derivations in particular physical cases~\cite{budkov2022modified,vasileva2023theory,budkov2023macroscopic}, it is instructive to discuss a general derivation. Let us set the origin of the $z$-axis at one wall, so that the other wall is located at $z=H$. We assume that the walls create the external field with potential energy $w(z)$. We can evaluate the disjoining pressure using the following thermodynamic relation~\cite{derjaguin1987derjaguin}:
\begin{equation}
\label{Pi_def}
\Pi=-\frac{\partial\Omega}{\partial H}-P_b,
\end{equation}
where in this case $\Omega$ denotes the GTP per unit area of the walls and $P_{b}$ is the pressure of the fluid in its bulk phase. We would like to point out that the differentiation of the GTP with respect to $H$ must be performed taking into account the fact that the order parameter satisfies the Euler-Lagrange equation ${\delta\Omega}/{\delta\psi(z)} = 0$ and that external potential, $w(z)$, depends on $H$. Let us assume that $\bar\psi(z)$ is the solution to the Euler-Lagrange equation, and $\psi_b = \bar\psi(H)$ is its boundary value. We assume that this boundary value is constant, when we take the derivative with respect to $H$ in equation (\ref{Pi_def}). For example, it could describe a situation where the electrostatic potential is fixed on the electrode surface, or where the concentration of molecules or ions is fixed at zero on the wall surface~\cite{vasileva2023theory}. The GTP density $\omega(z)=\omega(\bar{\psi}(z),\bar{\psi}'(z),w(z))$, with $\bar{\psi}'(z)=\partial_z\bar{\psi}(z)$.
Thus, using theorem of differentiation of the integral with respect to a parameter, we get
\begin{equation}
\frac{\partial \Omega}{\partial H}=\omega(H) +\int\limits_{0}^{H}\bigg{(}\frac{\partial\omega}{\partial\bar{\psi}}\frac{\partial\bar{\psi}}{\partial H}+
\frac{\partial\omega}{\partial\bar{\psi}'}\frac{\partial\bar{\psi}'}{\partial H}+\frac{\partial{\omega}}{\partial w}\frac{\partial w}{\partial H}\bigg{)}=
\omega(H) + \frac{\partial \omega}{\partial \bar{\psi}'} \left(\frac{\partial \bar{\psi}}{\partial H}\right)_{z}\bigg{|}_{0}^{H}+\int\limits_{0}^{H}dz n\frac{\partial w}{\partial H},
\end{equation}
where we have integrated by parts the second term in the integrand and used that ${\delta\Omega}/{\delta\bar{\psi}}= 0$; the notation $\left({\partial \bar{\psi}}/{\partial H} \right)_z$ means that the derivative is taken with respect to $H$, with $z$ being held constant. Taking into account that $\left({\partial \bar{\psi}}/{\partial H}\right)_{z}{|}_{z=0}=0$ and that 
\begin{equation}
\frac{\partial \psi_b}{\partial H}=\left(\frac{\partial \bar{\psi}}{\partial H}\right)_{z}\bigg{|}_{z=H}+\left(\frac{\partial \bar{\psi}}{\partial z}\right)_{H}\bigg{|}_{z=H}=0,
\end{equation}
i.e. that $\left({\partial \bar{\psi}}/{\partial H}\right)_{H}{|}_{z=H}=-\bar{\psi}'(H)$,
we eventually obtain
\begin{equation}
\frac{\partial \Omega}{\partial H}=\left(\omega-\frac{\partial \omega}{\partial \bar{\psi}'}\bar{\psi}'\right)\bigg{|}_{z=H}+\int\limits_{0}^{H}dz n\frac{\partial w}{\partial H} = \sigma_{zz}(H)+\int\limits_{0}^{H}dz n\frac{\partial w}{\partial H}.
\end{equation}
where $\sigma_{zz}(H)=(\omega-\bar{\psi}^{\prime}{\partial\omega}/{\partial\bar{\psi}^{\prime}})|_{z=H}$, is a normal stress at $z=H$.
Hence, we arrive at the expression
\begin{equation}
\Pi=-\int\limits_{0}^{H}dz\,n(z)\frac{\partial w(z)}{\partial H}-\sigma_{zz}(H)-P_b,
\end{equation}

Now let us consider the case of two identical walls creating the following total external potential
\begin{equation}
\label{u_}
w(z)=\phi(z)+\phi(H-z),
\end{equation}
where $\phi$ denotes the potential produced by one wall, and considering that $n(z)=n(H-z)$, we obtain
\begin{equation}
\label{Pi_}
\Pi=-\int\limits_{0}^{H}dz\,n(z)\phi^{\prime}(z)-\sigma_{zz}(H)-P_b.
\end{equation}
We can rewrite eq. (\ref{Pi_}) in a form more suitable for practical applications. Utilizing the mechanical equilibrium condition
\begin{equation}
\frac{d\sigma_{zz}(z)}{dz}-n(z)w^{\prime}(z)=0,
\end{equation}
after the integration from $z=H/2$ to $z=H$, we obtain
\begin{equation}
\label{sigmazz_}
\sigma_{zz}(H)=\sigma_{zz}\left(\frac{H}{2}\right)+\int\limits_{H/2}^{H}dz\,n(z)\phi^{\prime}(z)-\int\limits_{0}^{H/2}dz\,n(z)\phi^{\prime}(z),
\end{equation}
where we take into account equation (\ref{u_}) and the condition $n(z)=n(H-z)$. Substituting (\ref{sigmazz_}) for $\sigma_{zz}(H)$ into the equation (\ref{Pi_}), after simple algebraic transformations, we obtain
\begin{equation}
\label{Pi3}
\Pi=P_{n}-P_{b}-2\int\limits_{H/2}^{H}dz\,n(z)\phi^{\prime}(z),
\end{equation}
where $P_{n}=-\sigma_{zz}\left({H}/{2}\right)$ denotes the normal pressure at the midpoint of the pore. Note that this expression for disjoining pressure for simple fluids was obtained in \cite{henderson1986compressibility} within a different approach. For the ionic fluids with the GTP density, determined by (\ref{IF_MF}) and external potentials acting on the ions $w_{\alpha}(z)=\phi_{\alpha}(z)+\phi_{\alpha}(H-z)$ the electrostatic disjoining pressure takes the form~\cite{budkov2022modified}
\begin{equation}
\label{Pi3_}
\Pi=P_{n}-P_{b}-2\sum\limits_{\alpha}\int\limits_{H/2}^{H}dz\,\bar{n}_{\alpha}(z)\phi_{\alpha}^{\prime}(z),
\end{equation}
where $\bar{n}_{\alpha}(z)=\partial P/\partial\bar{\mu}_{\alpha}(z)=\partial \omega/\partial w_{\alpha}(z)$ is the local ionic concentration of ions $\alpha$.

\subsection{Stress tensor for thermodynamic potentials, containing second derivatives of the order parameters}

\subsubsection{Stress tensor}
In practice, one often works with functionals that contain a scalar order parameter with derivatives of higher order than the first one. However, in such cases, the approach outlined above typically results in an asymmetric tensor $\bar{\sigma}_{ik}$. Let us consider a scenario where the thermodynamic potential density depends on the second derivatives of the electrostatic potential
\begin{equation}
\omega(\bold{r})=\omega(\psi(\bold{r}),\partial \psi(\bold{r}),\partial\partial \psi(\bold{r})).
\end{equation}
Subjecting the system to a small deformation, and considering that the potential satisfies the Euler-Lagrange equation
\begin{equation}
\frac{\delta\Omega}{\delta\psi}=\frac{\partial{\omega}}{\partial\psi}-\partial_{i}\bigg(\frac{\partial{\omega}}{\partial(\partial_{i}\psi)}\bigg)
+\partial_{i}\partial_{j}\bigg(\frac{\partial{\omega}}{\partial(\partial_{i}\partial_j\psi)}\bigg)=0,
\end{equation}
after similar to previous, but more complicated calculations, we obtain
\begin{equation}
\delta\Omega = \int d\bold{r}\partial_{i}u_{k}\bar{\sigma}_{ik},
\end{equation}
where the tensor
\begin{equation}
\label{sigma_tens_2}
\bar{\sigma}_{ik}=\omega\delta_{ik}-\frac{\partial \omega}{\partial(\partial_{i}\psi)}\partial_{k}\psi+
\partial_{j}\bigg(\frac{\partial \omega}{\partial(\partial_i\partial_j\psi)}\bigg)\partial_{k}\psi
-\frac{\partial \omega}{\partial(\partial_i\partial_j\psi))}\partial(\partial_j\partial_k\psi)
\end{equation}
is not symmetric in general case. However, it is always possible to perform the symmetrization according to the following rule:
\begin{equation}
\sigma_{ik}=\bar{\sigma}_{ik}+\partial_{l}A_{ilk},
\end{equation}
where the third rank tensor, $A_{ilk}$, antisymmetric with respect to the indices $i$ and $l$, should be expressed via the first and second partial derivatives of the potential. This substitution does not violate the mechanical equilibrium conditions (\ref{partial_sigma}) due to the antisymmetry of $A_{ilk}$. Taking this into account, we have:
\begin{equation}
\delta\Omega =\int d\bold{r}\,\partial_{i}u_{k}\bar{\sigma}_{ik}=\int d\bold{r}\,\partial_{i}u_{k}(\bar{\sigma}_{ik}+\partial_{l}A_{ilk})=\int d\bold{r}\,u_{ik}\sigma_{ik},
\end{equation}
where we took into account that
\begin{equation}
\int d\bold{r}\,\partial_{i}u_{k}\partial_{l}A_{ilk}=-\int d\bold{r} \,u_{k}\partial_{i}\partial_{l}A_{ilk}=0.
\end{equation}
In the last step, we performed integration by parts and used the antisymmetry of the tensor $A_{ilk}$ with respect to the indices $i$ and $l$. Thus, we arrive at the same thermodynamic expression (\ref{sigma_therm})

\subsubsection{Direct symmetrization of stress tensor}
We can symmetrize the tensor $\bar{\sigma}_{ik}$ in different ways. The method discussed below will allow us to determine the explicit form of the tensor $A_{ilk}$. Let us utilize the fact that the GTP must be invariant not only under constant shift, but also under rotation of the system as a whole. In this subsection, for brevity, we will use the following notations for coordinate partial derivatives: $\partial_i\psi = \psi_{,i}$, and $\partial_{i}\partial_{k} \psi = \psi_{,ik}$. 

The transformation of infinitesimal rotation, as is known, has the form~\cite{weinberg1972gravitation}
\begin{equation}\label{}
{x}^{\prime}_{i}=x_{i}+\omega_{ij}x_{j},\quad \omega_{ij}=-\omega_{ji}=const.
\end{equation}
The condition of invariance of the GTP in this case results in the following relation
\begin{equation}
\label{torque}
\partial_{j}L_{jkl}=0,
\end{equation}
where we introduce the following torque tensor
\begin{equation}
\label{torque_2}
L_{jkl}=\bar{\sigma}_{jk}x_{l}-\bar{\sigma}_{jl}x_{k}+S_{kjl}-S_{ljk},
\end{equation}
where
\begin{equation}\label{spin}
S_{kjl}=\frac{\partial \omega}{\partial(\partial_j\partial_k\psi)}\partial_l\psi
\end{equation}
is the 'spin' part of the torque tensor.

Utilizing (\ref{torque}), we obtain
\begin{equation}\label{5}
\bar{\sigma}_{ik}-\bar{\sigma}_{ki}=2\partial_{l}\varphi_{ikl},
\end{equation}
where
\begin{equation}\label{varphi_1}
\varphi_{ikl}=\frac{1}{2}\bigg(S_{ilk}-S_{kli}\bigg).
\end{equation}
Thereby, we have
\begin{equation}\label{varphi_2}
\varphi_{ikl}=-\varphi_{kil}.
\end{equation}
Thus, we can construct the following symmetric stress tensor
\begin{equation}\label{MPP}
\sigma_{ik}=\frac{1}{2}\bigg(\bar{\sigma}_{ik}+\bar{\sigma}_{ki}\bigg)-\partial_{l}\bigg(\varphi_{ilk}+\varphi_{kli}\bigg),
\end{equation}
which leads to the following result obtained for the first time within the general covariant approach in~\cite{brandyshev2023noether}:
\begin{equation}\label{sigma_tens_4}
\sigma_{ik}=\omega\delta_{ik}-\frac{1}{2}\bigg(\frac{\partial \omega}{\partial\psi_{,i}}\psi_{,k}+\frac{\partial \omega}{\partial\psi_{,k}}\psi_{,i}\bigg)
-\partial_{j}\left(\frac{\partial \omega}{\partial\psi_{,ik}}\psi_{,j}\right)
+\partial_{j}\bigg(\frac{\partial \omega}{\partial\psi_{,ij}}\bigg)\psi_{,k}
+\partial_{j}\bigg(\frac{\partial \omega}{\partial\psi_{,kj}}\bigg)\psi_{,i}.
\end{equation}

Note that the structure of the tensor (\ref{MPP}) follows the general form of the stress tensor proposed by Martin, Parodi, and Pershan~\cite{martin1972unified,landau2012theory}. As can be seen from (\ref{MPP}), the simplest method of symmetrization, $(\sigma_{ik} + \sigma_{ki}) / 2$, will not produce a divergence-free tensor.

The condition (\ref{torque}) also leads us to the following tensor equality
\begin{equation}
\label{identity}
\frac{1}{2}\bigg(\frac{\partial \omega}{\partial\psi_{,i}}\psi_{,k}-\frac{\partial \omega}{\partial\psi_{,k}}\psi_{,i}\bigg)+
\frac{\partial \omega}{\partial\psi_{,ij}}\psi_{,jk}-\frac{\partial \omega}{\partial\psi_{,kj}}\psi_{,ji}=0,
\end{equation}
which is nothing more than the invariance condition of the GTP under rotations. With the help of this latter, the stress tensor (\ref{sigma_tens_4}) can be rewritten in a form that was also recently obtained using a general covariant approach~\cite{brandyshev2023noether}
\begin{multline}
\label{sigma_tens_5}
\sigma_{ik}=\omega\delta_{ik}-\frac{\partial \omega}{\partial\psi_{,i}}\psi_{,k}+
\partial_{j}\bigg(\frac{\partial \omega}{\partial\psi_{,ij}}\bigg)\psi_{,k}\\
-\partial_{j}\left(\frac{\partial \omega}{\partial\psi_{,ik}}\psi_{,j}\right)
+\partial_{j}\bigg(\frac{\partial \omega}{\partial\psi_{,jk}}\bigg)\psi_{,i}
+\frac{\partial \omega}{\partial\psi_{,kj}}\psi_{,ji}-\frac{\partial \omega}{\partial\psi_{,ij}}\psi_{,jk}.
\end{multline}
This result straightforwardly transforms into the expression (\ref{sigma_tens}), when the GTP density does not depend on the second derivatives. Moreover, formula (\ref{sigma_tens_5}) can be easily generalized to the case when the GTP depends on $m$ order parameters, $\psi^{(s)}$, $s=1,2,...,m$.

Note that when the GTP density does not depend on second derivatives, the condition of its invariance with respect to rotations gives
\begin{equation}
\label{identity}
\frac{\partial \omega}{\partial\psi_{,i}}\psi_{,k}=\frac{\partial \omega}{\partial\psi_{,k}}\psi_{,i},
\end{equation}
whence follows the symmetry of the tensor (\ref{sigma_tens}). From the expression (\ref{sigma_tens_5}) it is straightforward to obtain the explicit form of the function $A_{ilk}$:
\begin{equation}
A_{ilk}= S_{ilk}-S_{lik}.
\end{equation}

\subsubsection{Force and torque}
Note that the torque tensor can be used to calculate the macroscopic torque acting on a body immersed in a fluid using the following tensor relations:
\begin{equation}\label{}
L_i=\varepsilon_{ikj}L_{kj},
\end{equation}
\begin{equation}\label{}
L_{kj}=\oint df n_{l} L_{lkj},
\end{equation}
where the integration is performed over the surface of the body with $df$ being the surface element, $\varepsilon_{ikj}$ denotes the absolute antisymmetric Levi-Civita symbol, and $n_i$ is the normal vector constructed at the chosen point on the surface of the body. Note that the force acting on a body placed in the ionic fluid can be calculated using the formula
\begin{equation}
\label{force}
F_{i}= \oint \sigma_{ik}n_{k} df,
\end{equation}
It is worth noting that, in the absence of external forces acting fluid molecules, integration can be performed over any closed surface that contains a body immersed in an ionic fluid. However, when there are external forces acting, the force on the body must be calculated using a formula derived from the mechanical equilibrium of an arbitrary volume of fluid
\begin{equation}
\label{force2}
F_{i}=\oint\limits \sigma_{ik}n_{k} df +\int f_{i}d\bold{r},
\end{equation}
where $f_{i}=-\sum_{\alpha}{n}_{\alpha}{\partial u_{\alpha}}/{\partial x_{i}}$ are the components of the non-electrostatic forces density, where the integration in the second term is carried out over the fluid volume, enclosed between the surface of the immersed body and the surface, over which the first term is integrated.

\subsubsection{Example of application to ionic fluids}
In conclusion of this section, we apply the formalism described above to obtain the stress tensor in the framework of the Bazant-Storey-Kornyshev (BSC) ionic fluid model, which accounts for short-range ionic correlations at the phenomenological level. The thermodynamic potential density of the BSC model is given by:
\begin{equation}
\omega=-\frac{\varepsilon}{8\pi}\left(\left(\nabla\psi\right)^2+l_{c}^2\left(\nabla^2\psi\right)^2\right)+\rho\psi+f(\{n_{\alpha}\})-\sum\limits_{\alpha}\mu_{\alpha}n_{\alpha},
\end{equation}
where the first term, in contrast to the mean-field Poisson-Boltzmann model, contains an additional ''nonlocal'' term $\varepsilon l_c^2(\nabla^2\psi)^2/2$, which is associated with ionic correlations beyond the mean-field theory; $\rho=\sum_{\alpha}q_{\alpha}n_{\alpha}$ is the local charge density of ions, $q_{\alpha}$ is the charge of ion $\alpha$. The correlation length $l_c$ represents the only phenomenological parameter of the model responsible for incorporating correlations of the ions. While the original BSC model~\cite{bazant2011double} employs the symmetric lattice gas model as the reference fluid model, other models with different dependencies of the free energy density, $f(\{n_{\alpha}\})$, on the concentrations, $n_{\alpha}$, can be utilized for practical calculations. The Euler-Lagrange equations corresponding to this potential are as follows
\begin{equation}
\label{EL_}
\nabla^2\psi-l_c^2\Delta^2\psi = -\frac{4\pi}{\varepsilon}\sum\limits_{\alpha}q_{\alpha}n_{\alpha},~\frac{\partial{f}}{\partial{n_{\alpha}}}=\mu_{\alpha}-q_{\alpha}\psi,
\end{equation}
where $\mu_{\alpha}$ is the bulk chemical potential of $\alpha$th component. Using (\ref{sigma_tens_5}), we arrive at the following stress tensor
\begin{equation}
\sigma_{ik}=\sigma_{ik}^{(h)}+\sigma_{ik}^{(el)},
\end{equation}
where $\sigma_{ik}^{(h)}=-P\delta_{ik}$ is previously defined hydrostatic stress tensor, with $P=\sum_{\alpha}n_{\alpha}{\partial{f}}/{\partial{n_{\alpha}}}-f$ is the local osmotic pressure of ions. The electrostatic stress tensor in this case consists of two terms:
\begin{equation}
\sigma_{ik}^{(el)}=\sigma_{ik}^{(M)}+\sigma_{ik}^{(cor)},
\end{equation}
where $\sigma_{ik}^{(M)}$ is the Maxwell stress tensor and $\sigma_{ik}^{(cor)}$ is the stress tensor due to the correlation of ions, which has the following form:
\begin{equation}
\label{sigmacor}
\sigma_{ik}^{(cor)}=\frac{\varepsilon l_{c}^2}{4\pi}\left(\left(\bold{\mathcal{E}}\cdot\nabla^2\bold{\mathcal{E}}+\frac{1}{2}\left(\nabla\cdot\bold{\mathcal{E}}\right)^2\right)\delta_{ik}-\mathcal{E}_{i}\nabla^2 \mathcal{E}_{k}-\mathcal{E}_{k}\nabla^2 \mathcal{E}_{i}\right).
\end{equation}
For the first time, eq. (\ref{sigmacor}) was obtained in ref.~\cite{de2020interfacial} using a different approach.

\subsection{Stress tensor for fluids described by vector order parameters}

Now, let us examine another important case in the theory of inhomogeneous fluids: functionals defined on vector order parameters. Obtaining stress tensors for these cases is essential for modeling the forces that arise between solids and a dielectric fluid, which is intrinsically strongly structured. In this scenario, polarization serves as a vector order parameter~\cite{blossey2022comprehensive,blossey2022continuum,blossey2022field,hedley2023dramatic}. Thus, we will use the notation $\bold{P}(\bold{r})$ to denote such an order parameter. Note that such models can be used to describe the structure of the polar solvent in electrolyte solutions confined within charged pores, as well as the solvation forces between macroscopic bodies immersed in polar fluids. As above, in this subsection, for brevity, we will use the following notations for partial coordinate derivatives of the vector order parameter's components: $\partial_k P_{i}=P_{i,k}$ and $\partial_k\partial_l P_{i}=P_{i,kl}$.

Thus, the GTP in this case takes the form
\begin{equation}
\label{omega_}
\Omega =\int d\bold{r} \omega(\bold{P},\partial \bold{P},\partial\partial \bold{P}).
\end{equation}
For clarity, we assume that the GTP density is independent of scalar order parameters. However, in real systems, this assumption may not hold. The dependence on scalar order parameters can be easily addressed separately using the method described above.

As it was considered above, let us examine a translation on constant vector $\bold{h}$:
\begin{equation}
\label{}
x'_{i}=x_{i}+h_{i},
\end{equation}
Then, variation of the GTP is
\begin{equation}
\label{Omega}
\delta\Omega=\int d\bold{r} \bigg[\partial_{i}(\omega h_i)+\delta\omega\bigg],
\end{equation}
where the following notation is introduced
\begin{equation}
\label{partial}
\partial_{i}\omega=
\left(\frac{\partial \omega}{\partial x_i}\right)_{P}
+
\frac{\partial \omega}{\partial P_l}
P_{l,i}
+\frac{\partial \omega}{\partial P_{l,j}}
P_{l,ij}
+\frac{\partial \omega}{\partial P_{l,jk}}
P_{l,ijk}.
\end{equation}
The partial derivative $({\partial \omega}/{\partial x_i})_P$ is calculated as a derivative of an explicit function at $\bold{P} = const$. The variation of the GTP density due to the change of the order parameter is
\begin{multline}\label{}
\delta\omega=\bigg(\frac{\partial\omega}{\partial P_{i}}
-\partial_{j}\bigg(\frac{\partial\omega}{\partial P_{i,j}}\bigg)
+\partial_{k}\partial_{j}\bigg(\frac{\partial\omega}{\partial P_{i,kj}}\bigg)\bigg)\delta P_{i}\\
+\partial_{j}\bigg(\frac{\partial\omega}{\partial P_{i,j}}\delta P_{i}
-\partial_{k}\bigg(\frac{\partial\omega}{\partial P_{i,kj}}\bigg)\delta P_{i}
+\frac{\partial\omega}{\partial P_{i,kj}}\delta P_{i,k}\bigg),
\end{multline}
whereas the variation of the order parameter is
\begin{equation}\label{9}
\delta P_{i}=P_{i}{}'(\bold{r})-P_{i}(\bold{r})=-h_{j}P_{i,j}.
\end{equation}
The Euler-Lagrange equations for the vector parameter are
\begin{equation}\label{}
\frac{\delta \Omega}{\delta P_{i}}=\frac{\partial\omega}{\partial P_{i}}
-\partial_{j}\bigg(\frac{\partial\omega}{\partial P_{i,j}}\bigg)
+\partial_{k}\partial_{j}\bigg(\frac{\partial\omega}{\partial P_{i,kj}}\bigg)=0.
\end{equation}
Then, using eq. (\ref{Omega}), we obtain
\begin{equation}\label{Omega2}
\delta\Omega=h_{k}\int d\bold{r}\;\partial_{i}\bar{\sigma}_{ik}= h_{i}\int d\bold{r}\;\left(\frac{\partial \omega}{\partial x_i}\right)_{P},
\end{equation}
where
\begin{equation}
\label{6}
\bar{\sigma}_{ik}=\omega\delta_{ik}+j_{ik},
\end{equation}
with the following auxiliary tensor
\begin{equation}
\label{jp}
j_{ik}=-\frac{\partial \omega}{\partial P_{j,i}}P_{j,k}+
\partial_{n}\bigg(\frac{\partial \omega}{\partial P_{j,ni}}\bigg)P_{j,k}
-\frac{\partial \omega}{\partial P_{j,ni}}P_{j,nk},
\end{equation}
and we took into account eq. (\ref{partial}). In eq. (\ref{omega_}), the GTP density, $\omega$, does not explicitly depend on the coordinates $x_i$, so it follows from equation (\ref{Omega2}) that the GTP (\ref{omega_}) is invariant under translation, and the following equation holds:
\begin{equation}\label{3}
\partial_{i}\bar{\sigma}_{ik}=0.
\end{equation}
As with scalar order parameters, the tensor $\bar{\sigma}_{ik}$ is generally not symmetric. To make it symmetric, we can apply the technique formulated above, which is based on the invariance of the GTP under global rotations
\begin{equation}\label{}
x'_{i}=x_{i}+\omega_{ij}x_{j},\quad \omega_{ij}=-\omega_{ji}=const,
\end{equation}
so
\begin{equation}\label{8}
h_{i}=\omega_{ij}x_{j}.
\end{equation}
By substituting eq. (\ref{8}) into eq. (\ref{9}), we obtain the transformation of the vector order parameter as follows
\begin{equation}\label{1}
\delta P_{i}=\frac{1}{2}\omega_{kj}\bigg(\delta_{ik}P_{j}-\delta_{ij}P_{k}
+x_{k}\partial_{j}P_{i}-x_{j}\partial_{k}P_{i}\bigg).
\end{equation}
Thus, by substituting (\ref{1}) into (\ref{Omega}) and assuming that the variation of the thermodynamic potential is zero, we obtain
\begin{equation}\label{4}
\partial_{j}L_{jkn}=0,
\end{equation}
where the torque tensor is
\begin{equation}\label{M}
L_{jkn}=\bar{\sigma}_{jk}x_{n}-\bar{\sigma}_{jn}x_{k}+S_{kjn}-S_{njk},
\end{equation}
and its spin part is
\begin{equation}\label{S}
S_{kjn}=\frac{\partial \omega}{\partial P_{k,j}}P_{n}
-\partial_{m}\bigg(\frac{\partial \omega}{\partial P_{k,jm}}\bigg)P_{n}
+\frac{\partial \omega}{\partial P_{k,jm}}P_{n,m}
-\frac{\partial \omega}{\partial P_{i,nj}}P_{i,k}.
\end{equation}

Using eq. (\ref{3}), we can simplify  eq. (\ref{4}) to
\begin{equation}\label{5}
\bar{\sigma}_{ik}-\bar{\sigma}_{ki}=2\partial_{l}\varphi_{ikl},
\end{equation}
where
\begin{equation}
\label{7}
\varphi_{ikl}=\frac{1}{2}\bigg(S_{ilk}-S_{kli}\bigg).
\end{equation}
Therefore, we have
\begin{equation}\label{phi}
\varphi_{ikl}=-\varphi_{kil}.
\end{equation}
Thus, we can construct the following explicitly symmetric tensor
\begin{equation}\label{stress}
\sigma_{ik}=\frac{1}{2}\bigg(\bar{\sigma}_{ik}+\bar{\sigma}_{ki}\bigg)-\partial_{l}\bigg(\varphi_{ilk}+\varphi_{kli}\bigg).
\end{equation}
Further, using eq. (\ref{5}), we can show that
\begin{equation}\label{10}
\sigma_{ik}-\bar{\sigma}_{ik}=-\partial_{l}\bigg(\varphi_{ilk}+\varphi_{kli}-\varphi_{kil}\bigg),
\end{equation}
where, taking into account eq. (\ref{phi}), in the right-hand side of eq. (\ref{10}), we obtain the derivative of an expression that is antisymmetric with respect to $i$ and $l$. Therefore
\begin{equation}\label{}
\partial_{i}\partial_{l}\bigg(\varphi_{ilk}+\varphi_{kli}-\varphi_{kil}\bigg)\equiv0.
\end{equation}
Thus, tensor (\ref{stress}) is equivalent to tensor (\ref{6}), up to a divergence-free term. This equivalence implies that both tensors satisfy the local mechanical equilibrium condition. Therefore, the stress tensor, taking into account (\ref{7}), assumes an explicitly symmetric form
\begin{equation}\label{general_sigma}
\sigma_{ik}=\frac{1}{2}\bigg[\bar{\sigma}_{ik}+\bar{\sigma}_{ki}-\partial_{l}\bigg(S_{ikl}-S_{lki}+S_{kil}-S_{lik}\bigg)\bigg]
\end{equation}
that is also written in the Martin-Parodi-Pershan form.

We will not provide lengthy explicit expressions for the stress tensors of specific polar fluid models\cite{blossey2022field,blossey2022continuum,blossey2023comprehensive,hedley2023dramatic}, but rather, we will limit ourselves to a more general expression~(\ref{general_sigma}), which represents the new result of this study.

\section{Account of fluctuations of the electrostatic potential within variational field theory}

Previously, we explored the mean field theory of inhomogeneous fluids without considering the fluctuations of the order parameters near the mean-field configuration induced by the thermal motion of fluid particles. In this section, we will discuss the field theory with account of fluctuations. In contrast to the previous general considerations that are applicable to various fluid systems with scalar or vector order parameters, this section will focus on the variational field theory approach as applied to the Coulomb gas, which is a system of point-like charges embedded in a continuous dielectric medium. This approach is known as variational field theory~\cite{wang2010fluctuation_,lue2006variational_}. We will demonstrate how we can apply the TMA using the variational theory of ionic fluids, taking into account the thermal fluctuations of the electrostatic potential.

\subsection{Brief discussion of variational field theory}

The grand partition function can be expressed as the following functional integral over the fluctuating electrostatic potential $\varphi(\bold{r})$~\cite{wang2010fluctuation_,budkov2023variational}:
\begin{equation}
\label{Xi_}
\Xi = \int \frac{\mathcal{D}\varphi}{C_0}\exp\left[{-\frac{\beta}{2}\left(\varphi G_0^{-1}\varphi\right)+i(\rho_{ext}\varphi)+\sum\limits_{\alpha}\bar{z}_{\alpha}\int d\bold{r}\,e^{i\beta q_{\alpha}\varphi(\bold{r})-\beta{w_{\alpha}(\bold{r})}}}\right],
\end{equation}
where $\bar{z}_{\alpha}=z_{\alpha}e^{\beta q_{\alpha}^2G_{0}(0)/2}$ are the renormalized fugacities and $\rho_{ext}(\bold{r})$ is the density of external charges; $q_{\alpha}$ are charges of ions, $\beta=(k_{B}T)^{-1}$, $k_{B}$ is the Boltzmann constant, $T$ is the temperature. The fluctuating electrostatic potential satisfies the boundary conditions on the external body surfaces, where $-\varepsilon\partial_n\varphi=i\sigma$ represents the surface charge on the external body, and $\partial_n=\bold{n}\cdot \nabla$ is the normal derivative at a specific point on the body surface. 

We have also introduced the following notations
\begin{equation}
\left(\varphi G_0^{-1} \varphi\right)=\int d\bold{r}\int d\bold{r}'\,\varphi(\bold{r})G_0^{-1}(\bold{r},\bold{r}')\varphi(\bold{r}'),
\end{equation}
\begin{equation}
(\rho_{ext}\varphi)=\int d\bold{r}\,\rho_{ext}(\bold{r})\varphi(\bold{r}),
\end{equation}
\begin{equation}
C_0=\int\mathcal{D}\varphi \exp\left[-\frac{\beta}{2}(\varphi G_{0}^{-1}\varphi)\right].
\end{equation}
The 'bare' Green's function is
\begin{equation}
G_{0}(\bold{r}-\bold{r}^{\prime})=\frac{1}{\varepsilon|\bold{r}-\bold{r}^{\prime}|}.
\end{equation}
The inverse bare Green's function,
\begin{equation}
\label{inv_Green}
G_{0}^{-1}(\bold{r},\bold{r}^{\prime})=-\frac{\varepsilon}{4\pi}\nabla^2\delta(\bold{r}-\bold{r}^{\prime}),
\end{equation} 
satisfies the relation
\begin{equation}
\int d\bold{r}^{\prime\prime}\, G_{0}^{-1}(\bold{r},\bold{r}^{\prime\prime})G_{0}(\bold{r}^{\prime\prime}-\bold{r}^{\prime})=\delta(\bold{r}-\bold{r}^{\prime}).
\end{equation}

Note that term $(\rho_{ext}\varphi)$ can be excluded after integration by parts in term $(\varphi G_0^{-1}\varphi)/2$ and using aforementioned boundary conditions for the fluctuating electrostatic potential, $-\varepsilon \partial_n \varphi =i\sigma$. This leads to
\begin{equation}\label{}
\Xi=\int \frac{\mathcal{D}\varphi}{C_0} \exp\bigg[-\beta{\mathcal{H}[\varphi]}\bigg],
\end{equation}
where we have introduced the field-theoretic Hamiltonian of the Coulomb gas
\begin{equation}
\label{H}
\mathcal{H}[\varphi]=\frac{\varepsilon}{8\pi}\int d\bold{r}(\nabla\varphi(\bold{r}))^2-\int d\bold{r}\,\hat{P}(\bold{r}).
\end{equation}
with fluctuating osmotic pressure of the ions
\begin{equation}
\hat{P}(\bold{r})=k_{B}T\sum\limits_{\alpha}\bar{z}_{\alpha}e^{i\beta q_{\alpha}\varphi(\bold{r})-\beta{w_{\alpha}}}.
\end{equation}
In the next subsection, we will use this representation of the grand partition function. In this subsection, we will work with the functional integral (\ref{Xi_}), which contains the term $(\rho_{ext} \varphi)$ in the integrand exponent.

Let us briefly discuss the variational field theory formalism~\cite{wang2010fluctuation_,lue2006variational_,budkov2023variational}. By shifting the field variable $\varphi \rightarrow \varphi+\varphi_0$ and transitioning to the new Green's function $G_0 \rightarrow G$, we arrive at the following expression:
\begin{multline}
\Xi = \frac{C}{C_0}\int \frac{\mathcal{D}\varphi}{C}\exp\bigg{[}-\frac{\beta}{2}\left(\varphi G^{-1}\varphi\right)-\frac{\beta}{2}\left(\varphi [G_0^{-1}-G^{-1}]\varphi\right)\\
+i(\rho_{ext}(\varphi+\varphi_0))-\beta (\varphi G_0^{-1}\varphi_0)-\frac{\beta}{2}(\varphi_0G_{0}^{-1}\varphi_0)\\
+\sum\limits_{\alpha}\bar{z}_{\alpha}\int d\bold{r}\,e^{i\beta q_{\alpha}\varphi_0(\bold{r})+i\beta q_{\alpha}\varphi(\bold{r})-\beta{w_{\alpha}}}\bigg{]},
\end{multline}
where we have introduced the notation for the normalization constant of the new Gaussian measure
\begin{equation}
C=\int {\mathcal{D}\varphi}\,\exp\bigg{[}-\frac{\beta}{2}\left(\varphi G^{-1}\varphi\right)\bigg{]}.
\end{equation}
Then, applying the Bogolyubov's inequality,
\begin{equation}
\left<e^{X}\right>\geq e^{\left<X\right>},
\end{equation}
to the functional integral, we obtain
\begin{equation}
\Xi \geq \exp W[G;\varphi_0],
\end{equation}
where we have introduced the following auxiliary functional
\begin{multline}
\label{W}
W[G;\varphi_0]= \frac{1}{2}\ln \frac{Det G}{Det G_0}-\frac{1}{2}tr\left(G [G_0^{-1}-G^{-1}]\right)+i(\rho_{ext}\varphi_0)-\frac{\beta}{2}(\varphi_0G_{0}^{-1}\varphi_0)\\
+\sum\limits_{\alpha}{z}_{\alpha}\int d\bold{r}\,e^{i\beta q_{\alpha}\varphi_0(\bold{r})+\frac{\beta q_{\alpha}^2}{2}(G_{0}(0)-G(\bold{r},\bold{r}))-\beta{w_{\alpha}}},
\end{multline}
where $Det G$ and $Det G_0$ are the functional determinants of corresponding integral operators~\cite{budkov2023variational} and $tr A$ is the trace of the integral operator in accordance with definition
\begin{equation}
tr A =\int d\bold{r} A(\bold{r},\bold{r}).
\end{equation}

The functions $\varphi_0=i\psi$ and $G(\bold{r},\bold{r}')$ are determined from the Euler-Lagrange equations for the variational functional (\ref{W}) (see also~\cite{budkov2023variational})
\begin{equation}
\label{EL_eqs}
\frac{\delta W}{\delta \varphi_0(\bold{r})}\bigg{|}_{\varphi_{0}=i\psi}=0, \quad \frac{\delta W}{\delta G(\bold{r},\bold{r}')}=0.
\end{equation}
The first one takes the following form
\begin{equation}
\label{psi_}
\nabla^2 \psi(\bold{r})=-\frac{4\pi}{\varepsilon}\left(\sum\limits_{\alpha} q_{\alpha} z_{\alpha} A_{\alpha}(\bold{r})e^{-\beta q_{\alpha}\psi(\bold{r})-\beta{w_{\alpha}}}+\rho_{ext}(\bold{r})\right),
\end{equation}
where
\begin{equation}
A_{\alpha}(\bold{r})=\exp\left[\frac{\beta q_{\alpha}^2}{2}\left(G_0(0)-G(\bold{r},\bold{r})\right)\right].   
\end{equation}
The second one takes the form of
\begin{equation}
\label{G-1_ionic_gas}
G^{-1}(\bold{r},\bold{r}')-G_0^{-1}(\bold{r},\bold{r}')=\beta \sum \limits_{\alpha} q_{\alpha}^2 z_{\alpha} A_{\alpha}(\bold{r})e^{-\beta q_{\alpha}\psi(\bold{r})-\beta{w_{\alpha}}}\delta(\bold{r}-\bold{r}').
\end{equation}
Using eq. (\ref{inv_Green}) and inverting the operator $G^{-1}$, eq. (\ref{G-1_ionic_gas}) can be rewritten as~\cite{budkov2023variational}
\begin{equation}
\label{Green_func_}
\left(- \nabla^2+\varkappa^2(\bold{r})\right)G(\bold{r},\bold{r}')=\frac{4\pi}{\varepsilon}\delta(\bold{r}-\bold{r}'),
\end{equation}
where
\begin{equation}
\varkappa^2(\bold{r})=\frac{4\pi}{\varepsilon k_{B}T}\sum\limits_{\alpha}q_{\alpha}^2 n_{\alpha}(\bold{r})
\end{equation}
and we have introduced the notations for the local concentrations of the ions
\begin{equation}
\label{chem_eq_ion}
n_{\alpha}(\bold{r})=\frac{\delta \Omega}{\delta u_{\alpha}(\bold{r})}=z_{\alpha}e^{-\beta q_{\alpha}\psi(\bold{r})+\frac{\beta q_{\alpha}^2}{2}(G_{0}(0)-G(\bold{r},\bold{r}))-\beta{w_{\alpha}}}.
\end{equation}

Thus, the GTP of the Coulomb gas can be calculated using the formula~\cite{wang2010fluctuation_}:
\begin{multline}
\label{Omega_ion_gas}
\Omega = -k_{B}T W[G;i\psi] = \int d\bold{r}\left(-\frac{\varepsilon (\nabla \psi)^2}{8\pi}+\rho_{ext}\psi\right)+k_{B}T\int d\bold{r}\,(n_{+}+n_{-})\\ 
+\int d\bold{r} \,\mathcal{I}(\bold{r})\int\limits_{0}^{1} d\tau\left(G(\bold{r},\bold{r};\tau)-G(\bold{r},\bold{r})\right),
\end{multline}
where 
\begin{equation}
\mathcal{I}(\bold{r})=\frac{1}{2}\sum\limits_{\alpha}q_{\alpha}^2 n_{\alpha}(\bold{r}).
\end{equation}

The equations (\ref{chem_eq_ion}) define the chemical equilibrium conditions for ions, which can be rewritten in terms of the equality of chemical potentials
\begin{equation}
\mu_{\alpha} = \bar{\mu}_{\alpha} +q_{\alpha}\psi+u_{\alpha},
\end{equation}
where the intrinsic chemical potentials take the form
\begin{equation}
\bar{\mu}_{\alpha}= k_{B}T\ln (n_{\alpha}\Lambda_\alpha^3)+\frac{q_{\alpha}^2}{2}\left(G_{0}(0)-G(\bold{r},\bold{r})\right).
\end{equation}
The intermediate Green's function $G(\bold{r},\bold{r}';\tau)$ satisfies the equation
\begin{equation}
\label{Green_func_tau}
\left(- \nabla^2+\tau\varkappa^2(\bold{r})\right)G(\bold{r},\bold{r}';\tau)=\frac{4\pi}{\varepsilon}\delta(\bold{r}-\bold{r}').
\end{equation}

Knowledge of the solution to the Euler-Lagrange equations (\ref{EL_eqs}) allows us to use the averaging approximation
\begin{equation}
\label{average_approx}
\frac{1}{\Xi}\int \frac{\mathcal{D}\varphi}{C_0}\exp\bigg[-\beta\mathcal{H}[\varphi]\bigg]F[\varphi]\approx\int\frac{\mathcal{D}\eta}{C}\exp\bigg[-\frac{\beta}{2}(\eta G^{-1}\eta)\bigg]F[\varphi_0+\eta].
\end{equation}
The approximation (\ref{average_approx}) is due to the fact that, within the framework of variational field theory, the initial functional in the integrand exponent is approximated by a quadratic functional with the Green function, $G(\bold{r},\bold{r}')$, and the extremum $\varphi_0(\bold{r})$ determined from the Euler-Lagrange equations (\ref{EL_eqs}). In the next subsection, we will use eq. (\ref{average_approx}).

\subsection{Correlation stress tensor of Coulomb gas}

Now we would like to formulate a generalization of the method for obtaining stress tensor in ionic fluids, considering fluctuations of the electrostatic potential. Let us start from expressing the previously determined grand partition function of the Coulomb gas:
\begin{equation}\label{}
\Xi=\int \frac{\mathcal{D}\varphi}{C_0} \exp\bigg[-\beta{\mathcal{H}[\varphi]}\bigg],
\end{equation}
where $\mathcal{H}$ is the previously determined field-theoretic Hamiltonian by eq. (\ref{H}).

First, we consider the case when external potentials are zero, i.e. $w_{\alpha}=0$. Additionally, we assume that the external charges are surface charges, and their influence is incorporated through the boundary conditions of the self-consistent field equations. Similar to discussed above mean field theory case, we perform a virtual displacement of the system by an infinitesimal vector $\bold{h}$, i.e.
\begin{equation}\label{}
x'_{k}=x_{k}+h_{k},
\end{equation}
and require the invariance of the grand partition function, i.e.
\begin{equation}\label{}
\delta\ln\Xi=-\beta\bigg\langle\delta\mathcal{H}[\varphi]\bigg\rangle=0.
\end{equation}
Thus, after straightforward transformations, we obtain
\begin{equation}\label{}
\partial_{i}\bigg\langle\hat{\sigma}_{ik}h_{k}\bigg\rangle=\bigg\langle \frac{\delta\mathcal{H}[\varphi]}{\delta\varphi}h_{k}\partial_{k}\varphi \bigg\rangle,
\end{equation}
where we have introduced a stress tensor, defined on the fluctuating electrostatic field
\begin{equation}\label{}
\hat{\sigma}_{ik}=
\frac{\varepsilon}{8\pi}\delta_{ik}\partial_j\varphi\partial_j\varphi
-\frac{\varepsilon}{4\pi}\partial_i\varphi\partial_k\varphi-\hat{P}\delta_{ik}.
\end{equation}
Due to the arbitrariness of $h_{k}$ we obtain
\begin{equation}\label{}
\partial_{i}\langle\hat{\sigma}_{ik}\rangle=\bigg\langle\frac{\delta\mathcal{H}[\varphi]}{\delta\varphi}\partial_{k}\varphi\bigg\rangle,
\end{equation}
where averaging is understood in the usual sense:
\begin{equation}
\left<(\dots)\right>=\frac{1}{\Xi}\int \frac{\mathcal{D}\varphi}{ C_0}\exp\bigg(-\beta {\mathcal{H}[\varphi]}\bigg)(\dots).
\end{equation}
Further, taking into account the identity \cite{buyukdagli2020schwinger,brandyshev2023statistical}
\begin{equation}
\bigg\langle\frac{\delta\mathcal{H}[\varphi]}{\delta\varphi(\bold{r})}\partial_{k}\varphi(\bold{r})\bigg\rangle= \partial_{k}\frac{\delta\varphi(\bold{r})}{\delta\varphi(\bold{r})}=\partial_{k}\delta(\bold{0})=0,
\end{equation}
we arrive at the condition of mechanical equilibrium of ionic fluid taking into account thermal fluctuations of electrostatic potential
\begin{equation}
\label{Noether}
\partial_{i}\langle\hat{\sigma}_{ik}\rangle=0.
\end{equation}

In the presence of external fields, when $w_{\alpha}(\mathbf{r}) \neq 0$, the same calculations that were performed to derive equation (\ref{partial_sigma}) lead to the mechanical equilibrium condition,
\begin{equation}
\partial_{i}\langle\hat{\sigma}_{ik}\rangle-\sum\limits_{\alpha}\langle\hat{n}_{\alpha}\rangle\partial_{k}w_{\alpha}=0,
\end{equation}
where
\begin{equation}
\hat{n}_{\alpha}=-\frac{\partial \hat{P}}{\partial w_{\alpha}}=\bar{z}_{\alpha}e^{i\beta q_{\alpha}\varphi(\bold{r})-\beta{w_{\alpha}}}
\end{equation}
are fluctuating concentrations of the ions.

Let us calculate the stress tensor within the framework of the variational field theory formulated above. For this purpose, let us represent the field variable as $\varphi(\bold{r})=i\psi(\bold{r})+\eta(\bold{r})$, where $\psi(\bold{r})$ is the solution of the equation (\ref{psi_}), and $\eta$ is the fluctuation over which we need to integrate. Thus, we obtain
\begin{equation}
\sigma_{ik}=-P\delta_{ik}+\sigma_{ik}^{(M)}+\sigma_{ik}^{(cor)},
\end{equation}
where $\sigma_{ik}^{(M)}$ is already introduced Maxwell stress tensor (see eq. (\ref{sigma_ik_2})), $P=\left<\hat{P}\right>$ is the average osmotic pressure of the ions, and
\begin{equation}
\label{sigma_cor}
\sigma_{ik}^{(cor)}=\frac{\varepsilon}{8\pi}\delta_{ik}\left<\partial_j\eta\partial_j\eta\right>
-\frac{\varepsilon}{4\pi}\left<\partial_i\eta\partial_k\eta\right>
\end{equation}
is the correlation stress tensor, determined by correlations of fluctuations of electrostatic field components. Using the averaging approximation (\ref{average_approx}), performing functional integration with the use of the Wick's theorem, we arrive at the following correlation stress tensor~\cite{budkov2023variational}
\begin{equation}
\sigma_{ik}^{(corr)}=\frac{\varepsilon}{4\pi}\left(\frac{1}{2}\mathcal{D}_{ll}(\bold{r})\delta_{ik}-\mathcal{D}_{ik}(\bold{r})\right),
\end{equation}
where we introduce tensor
\begin{equation}
\mathcal{D}_{ik}(\bold{r})=\lim\limits_{\bold{r}^{\prime}\to\bold{r}}\partial_{i}\partial_{k}^{\prime}G(\bold{r},\bold{r}^{\prime})
\end{equation}
which is simply the autocorrelation function of the fluctuating electrostatic field.

For practical purposes, it is necessary to utilize the regularized tensor 
\begin{equation}
\mathcal{D}_{ik}(\bold{r})=\lim\limits_{\bold{r}^{\prime}\to\bold{r}}\partial_{i}\partial_{k}^{\prime}\bar{G}(\bold{r},\bold{r}^{\prime}),
\end{equation}
where 
\begin{equation}
\bar{G}(\bold{r},\bold{r}^{\prime})=G(\bold{r},\bold{r}^{\prime})-G_{0}(\bold{r}-\bold{r}^{\prime}).
\end{equation}
The regularization is performed to exclude singularity at $\bold{r}'=\bold{r}$. Note that subtracting $G_{0}(\mathbf{r} - \mathbf{r}^{\prime})$ does not change the mechanical equilibrium equation, $\partial_{i}\sigma_{ik} = 0$, because it results in the appearance of a divergence-free term in the total stress tensor~\cite{budkov2023variational,brandyshev2023statistical}.

The osmotic pressure of ions, according to the averaging rule (\ref{average_approx}), is as follows:
\begin{equation}
P=k_{B}T\sum\limits_{\alpha}\bar{z}_{\alpha}e^{-\frac{\beta q_{\alpha}^2}{2}G(\bold{r},\bold{r})-\beta q_{\alpha}\psi(\bold{r})-\beta{w_{\alpha}}}.
\end{equation}

\subsection{Interaction of charged conductive and dielectric plates in electrolyte solutions}
Let us examine the Coulomb gas confined within a slit-like pore with flat charged infinite walls. Let us assume that these walls generate external potentials
\begin{equation}
w_{\alpha}(z)=\phi_{\alpha}(z)+\phi_{\alpha}(H-z),
\end{equation}
where $\phi_{\alpha}(z)$ represents the potential of the external field acting on the $\alpha$-th ion at the point $z\in [0,H]$. The identical walls are situated at $z=0$ and $z=H$. The disjoining pressure is given by the expression
\begin{equation}
\Pi = -\frac{\partial (\Omega/\mathcal{A})}{\partial H}-P_{b},
\end{equation}
where $\mathcal{A}$ represents the total area of the walls' surface. By considering that solutions of self-consistent field equations (\ref{psi_}) and (\ref{G-1_ionic_gas}) minimize the functional $\Omega = -k_{B}T W[G;\psi]$, we obtain
\begin{equation}
\Pi = -\sum\limits_{\alpha}\int\limits_{-\infty}^{\infty}dz\,n_{\alpha}(z)\phi^{\prime}_{\alpha}(z)-P_{b},
\end{equation}
where we utilized the relation $n_{\alpha}(H-z)=n_{\alpha}(z)$, which follows from the symmetry of the problem, i.e., the identity of the pore walls. Integrating the mechanical equilibrium condition,
\begin{equation}
\label{eq_cond_2}
\frac{d \sigma_{zz}}{dz}- \sum\limits_{\alpha}n_{\alpha}(z)w_{\alpha}'(z)=0,
\end{equation}
from $z=H/2$ to $z=\infty$ and considering the condition $\sigma_{zz}(\infty)=0$, we arrive at
\begin{equation}
-\sigma_{zz}(H/2)=\sum\limits_{\alpha}\int\limits_{H/2}^{\infty} d z\, n_{\alpha}(z)w_{\alpha}'(z)=\sum\limits_{\alpha}\int\limits_{H/2}^{\infty} d z\, n_{\alpha}(z)\phi_{\alpha}'(z)-\sum\limits_{\alpha}\int\limits_{-\infty}^{H/2} d z\, n_{\alpha}(z)\phi_{\alpha}'(z).
\end{equation}
Therefore,
\begin{multline}
\Pi=-\sum\limits_{\alpha}\int\limits_{-\infty}^{\infty}dz\,n_{\alpha}(z)\phi^{\prime}_{\alpha}(z)-P_{b}=
-\sum\limits_{\alpha}\int\limits_{-\infty}^{H/2}dz\,n_{\alpha}(z)\phi^{\prime}_{\alpha}(z)-\sum\limits_{\alpha}\int\limits_{H/2}^{\infty}dz\,n_{\alpha}(z)\phi^{\prime}_{\alpha}(z)-P_b
\\=-\sigma_{zz}(H/2)-P_b-2\sum\limits_{\alpha}\int\limits_{H/2}^{\infty}dz\,n_{\alpha}(z)\phi_{\alpha}^{\prime}(z).
\end{multline}
Thus, given that $n_{\alpha}(z)=0$ at $z \geq H$ (impermeable wall), we finally obtain already known expression (see eq. (\ref{Pi3_}))
\begin{equation}
\label{disj_press}
\Pi=P_n-P_b-2\sum\limits_{\alpha}\int\limits_{H/2}^{H}dz\,n_{\alpha}(z)\phi_{\alpha}^{\prime}(z),
\end{equation}
where $P_{n}=-\sigma_{zz}(H/2)$ as above is the normal pressure in the pore midpoint.

Now let us specify the normal pressure in the middle of the pore. For a slit pore geometry, the Green's function can be expressed as
\begin{equation}
G(\bold{r},\bold{r}')=G(\boldsymbol{\rho}-\boldsymbol{\rho}'|z,z'),
\end{equation}
where $\boldsymbol{\rho}$ is a two-dimensional vector lying in the plane parallel to the pore walls, and the $z$-axis is perpendicular to the pore wall. Let us examine the two-dimensional Fourier transform:
\begin{equation}
\label{}
G(\boldsymbol{\rho}-\boldsymbol{\rho}'|z,z')=\int \frac{d^2\bold{p}}{(2\pi)^2}e^{-i\bold{p}(\boldsymbol{\rho}-\boldsymbol{\rho}')}G(p|z,z'),
\end{equation}
where $G(p|z,z')$ is an even function of $\bold{p}$, depending only on the absolute value $p=|\bold{p}|$. Nondiagonal elements of the tensor $\mathcal{D}_{ik}(z)$ are
\begin{equation}\label{}
\mathcal{D}_{xy}(z)=\mathcal{D}_{xz}(z)=\mathcal{D}_{yz}(z)=0,
\end{equation}
whereas the diagonal ones have the form
\begin{equation}\label{}
\mathcal{D}_{xx}(z)=\int \frac{d^2\bold{p}}{(2\pi)^2}p_x^2Q(q,z), \quad \mathcal{D}_{yy}(z)=\int \frac{d^2\bold{p}}{(2\pi)^2}p_y^2Q(q,z), \quad \mathcal{D}_{zz}(z)=\int \frac{d^2\bold{p}}{(2\pi)^2}\mathcal{D}_{zz}(q,z),
\end{equation}
where
\begin{equation}\label{Dzz}
\mathcal{D}_{zz}(p,z)=k_{B}T\lim_{z'\rightarrow z}\partial_z\partial_{z'}G(p|z,z'),
\end{equation}
\begin{equation}\label{Q}
Q(p,z)=k_{B}T\lim_{z'\rightarrow z}G(p|z,z'),
\end{equation}
so that
\begin{equation}
\mathcal{D}_{ll}(z)=\int \frac{d^2\bold{p}}{(2\pi)^2}\bigg(p^2Q(p,z)+\mathcal{D}_{zz}(p,z)\bigg).
\end{equation}
The Fourier transform of the Green's function, $G(p|z,z')$, can be found from the following equation
\begin{equation}\label{}
\frac{\varepsilon}{4\pi} \bigg(-\partial^2_z+p^2+\varkappa^2(z)\bigg)G(p|z,z')=\delta(z-z').
\end{equation}
Hence, the normal pressure in (\ref{disj_press}) has the following form
\begin{equation}
\label{Pn}
P_{n}=P_{m}+\frac{\varepsilon}{4\pi} \left(\mathcal{D}_{zz}\left(\frac{H}{2}\right)-\frac{1}{2}\mathcal{D}_{ll}\left(\frac{H}{2}\right)\right),
\end{equation}
where $P_{m}=P_0(H/2)$ denotes the osmotic pressure of the ions at the midpoint of the pore, and we have also assumed that $\psi'(H/2)=0$.

Finally, let us estimate the asymptotics of the second term on the right-hand side of equation (\ref{Pn}) corresponding to the electrostatic field fluctuations as $H\to \infty$. In this asymptotic regime, we have $\varkappa(z)\simeq\kappa$, and since
\begin{equation}
G(p|z,z')\simeq\frac{2\pi e^{-\sqrt{p^2+\kappa^2}|z-z'|}}{\varepsilon \sqrt{p^2+\kappa^2}}, \quad G_0(p|z,z')=\frac{2\pi e^{-p|z-z'|}}{\varepsilon p},
\end{equation}
we obtain~\cite{budkov2023variational}
\begin{multline}
\frac{\varepsilon}{4\pi} \left(\mathcal{D}_{zz}\left(\frac{H}{2}\right)-\frac{1}{2}\mathcal{D}_{ll}\left(\frac{H}{2}\right)\right)\\\simeq\frac{k_{B}T}{8\pi}\int\limits_{0}^{\infty}dp p\left(2p-\sqrt{p^2+\kappa^2}-\frac{p^2}{\sqrt{p^2+\kappa^2}}\right)=-\frac{k_B T \kappa^3}{24\pi},
\end{multline}
that is, the Debye-H\"{u}ckel expression, where $\kappa$ is the inverse Debye length. Note that the Debye-H\"{u}ckel result for the excess osmotic pressure of Coulomb gas additionally indicates the consistency of the presented formalism.

Let us now calculate the disjoining pressure between charged identical walls located at a sufficiently large distance from each other, such that we can neglect the contribution of specific interactions of ions with the walls. We can express it as
\begin{equation}
\label{Pi2}
\Pi = P_{n}-P_{b}=P_{m}+\frac{\varepsilon}{4\pi} \left(\mathcal{D}_{zz}\left(\frac{H}{2}\right)-\frac{1}{2}\mathcal{D}_{ll}\left(\frac{H}{2}\right)\right)-P_{b}.
\end{equation}
Let us estimate the asymptotic behavior of the disjoining pressure as $H\to \infty$, where $\varkappa(z)\simeq \kappa$. In this case, the Green's function is defined by the following equation:
\begin{equation}
\label{}
\frac{\varepsilon}{4\pi}\bigg(-\partial^2_z+p^2+\kappa^2\bigg)G(p|z,z')=\delta(z-z').
\end{equation}

\textbf{Metal walls.} First, let us consider the case of metal walls (the case of dielectric walls will be considered below), for which the boundary conditions are defined as follows
\begin{equation}
\begin{gathered}
\label{bound_metal}
G(p|0,z')=G(p|H,z')=0, \\ 
\frac{\varepsilon}{4\pi}\left(\frac{\partial G(p|z'-0,z')}{\partial z}-\frac{\partial G(p|z'+0,z')}{\partial z}\right)=1,\\ 
G(p|z'-0,z')=G(p|z'+0,z'),~z'\in (0;H).
\end{gathered}
\end{equation}
Such boundary conditions can be explained by the fact that the Green's function can be interpreted as the potential created by unit test point-like charge placed at point $\bold{r}'$.

The solution is as follows
\begin{equation}
\label{G}
G(p|z,z')=
\begin{cases}
\frac{4\pi\sinh \kappa_{p}z\left(\cosh \kappa_p z'-\coth\kappa_p H\sinh \kappa_p z'\right)}{\varepsilon \kappa_p}, &z<z',\\
\frac{4\pi\sinh \kappa_{p}z'\left(\cosh \kappa_p z-\coth\kappa_p H\sinh \kappa_p z\right)}{\varepsilon \kappa_p}, &z>z',
\end{cases}
\end{equation}
where $\kappa_p=\sqrt{p^2+\kappa^2}$. Normal pressure in the midpoint of the pore can be determined by the expression
\begin{multline}
\label{corr}
\frac{\varepsilon}{4\pi} \left(\mathcal{D}_{zz}\left(\frac{H}{2}\right)-\frac{1}{2}\mathcal{D}_{ll}\left(\frac{H}{2}\right)\right)=\frac{k_{B}T}{4}\int \frac{d^2\bold{p}}{(2\pi)^2} \left(2p-\frac{\kappa_p}{\tanh \frac{\kappa_p H}{2}}-\frac{p^2\tanh \frac{\kappa_p H}{2}}{\kappa_p}\right)\\
=\frac{k_{B}T}{4}\int\frac{d^2\bold{p}}{(2\pi)^2}\left(2p-\sqrt{p^2+\kappa^2}-\frac{p^2}{\sqrt{p^2+\kappa^2}}\right)\\
-\frac{k_{B}T}{4}\int\frac{d^2\bold{p}}{(2\pi)^2}\left(\kappa_p\left(\coth\frac{\kappa_p H}{2} -1\right)+\frac{p^2}{\kappa_p}\left(\tanh\frac{\kappa_p H}{2} -1\right)\right).
\end{multline}
Furthermore,
\begin{equation}
\label{Pn3}
P_{m}\simeq P_{0}+\frac{\varepsilon \kappa^2\psi_{m}^2}{8\pi}+\delta P_m,
\end{equation}
where $P_0=k_{B}T\sum_{\alpha} n_{\alpha}$ is the bulk pressure without taking into account the electrostatic interactions of the ions, $\delta P_n$ is its perturbation, arising from the electrostatic field fluctuations, and $\psi_{m}=\psi(H/2)$
\begin{equation}
\label{Pcor}
\delta P_m=-\frac{1}{2}\sum\limits_{\alpha}q_{\alpha}^2 n_{\alpha}\int \frac{d^2\bold{p}}{(2\pi)^2}\delta G\left(p\bigg{|}\frac{H}{2},\frac{H}{2}\right).
\end{equation}
We also introduce the notation
\begin{equation}
\delta G\left(p\bigg{|}\frac{H}{2},\frac{H}{2}\right)=G\left(q\bigg{|}\frac{H}{2},\frac{H}{2}\right)-\lim\limits_{H\to \infty}G\left(p\bigg{|}\frac{H}{2},\frac{H}{2}\right)=\frac{2\pi(\tanh\frac{\kappa_p H}{2}-1)}{\varepsilon \kappa_p}
\end{equation}
and take into account that $n_{\alpha}=\partial{P}/\partial{\mu_{\alpha}}$. Consequently, considering (\ref{corr}) and 
\begin{equation}
P_{b}=P_0 -\frac{k_{B}T\kappa^3}{24\pi},
\end{equation}
and substituting (\ref{Pn3}) into the equation (\ref{Pi2}), we obtain
\begin{equation}
\Pi =\Pi_{MF}+\Pi_{cor},
\end{equation}
where the mean-field contribution to the disjoining pressure is determined from the solution of linearized Poisson-Boltzmann equation through the well-known expression~\cite{barrat2003basic} 
\begin{equation}
\label{disj_press_MF}
\Pi_{MF}= \frac{2\pi \sigma^2}{\varepsilon} \frac{1}{\sinh^2\left(\frac{\kappa H}{2}\right)}\simeq \frac{8\pi \sigma^2}{\varepsilon} e^{-\kappa H},
\end{equation}
whereas the correlation contribution to the disjoining pressure takes the form:
\begin{equation}
\label{Pi_corr_metal}
\Pi_{cor} = -\frac{k_{B}T}{2}\int\frac{d^2\bold{p}}{(2\pi)^2}\kappa_p\left(\coth \kappa_{p}H-1\right)=-\frac{k_{B}T}{2\pi}\int\limits_{0}^{\infty}dp\,p\frac{\kappa_p}{e^{2\kappa_p H}-1}.
\end{equation}

The latter expression yields the following limiting cases
\begin{equation}
\label{disj_press_asymp}
\Pi_{cor}\simeq
\begin{cases}
-\frac{k_B T}{8\pi H^3}\zeta(3), & \kappa H\ll 1\,\\
-\frac{k_B T\kappa^2}{4\pi H}e^{-2\kappa H},&\kappa H\gg 1,
\end{cases}
\end{equation}
where $\zeta(x)$ is the Riemann zeta function, which have been obtained for the first time by Ninham and Yaminsky within a different method~\cite{ninham1997ion}. This result have been also obtained more recently within the variational field theory in~\cite{hatlo2008role}. Note that the contribution of the correlation to the disjoining pressure includes the contribution of the classical Casimir force, which is caused by thermal fluctuations in the electrostatic field~\cite{hatlo2008role,ninham1997ion,jancovici2004screening,dzyaloshinskii1961general}.

\textbf{Dielectric walls.}
Now let us consider the case of two identical dielectric walls with dielectric constant $\varepsilon_{w}$. In this case, the boundary conditions (\ref{bound_metal}) should be replaced by the following ones
\begin{equation}
\begin{gathered}
\label{bound_diel}
\varepsilon\frac{\partial G(p|+0,z')}{\partial z}=\varepsilon_{w}\frac{\partial G(p|-0,z')}{\partial z},\\
\varepsilon\frac{\partial G(p|H-0,z')}{\partial z}=\varepsilon_{w}\frac{\partial G(p|H+0,z')}{\partial z},\\
\frac{\varepsilon}{4\pi}\left(\frac{\partial G(p|z'-0,z')}{\partial z}-\frac{\partial G(p|z'+0,z')}{\partial z}\right)=1,\\ 
G(p|z'-0,z')=G(p|z'+0,z'),~z'\in (0;H),
\end{gathered}
\end{equation}
which correspond to the boundary conditions for the electrostatic potential on the interface of two dielectrics.

Taking into account these boundary conditions, the Green's function has the form
\begin{equation}
\label{G}
G(p|z,z')=
\begin{cases}
\frac{4\pi(\kappa_p\varepsilon\cosh\kappa_{p}(H-z^{\prime})+\varepsilon_{w} p\sinh\kappa_p(H-z^{\prime}))(\kappa_p\varepsilon\cosh\kappa_{p}z+\varepsilon_{w} p\sinh\kappa_pz)}{\varepsilon \kappa_p ((\kappa_p^2\varepsilon^2+\varepsilon_{w}^2p^2)\sinh\kappa_pH + 2\kappa_p p\varepsilon\varepsilon_{w}\cosh\kappa_pH)}, &z<z',\\
\frac{4\pi (\kappa_p\varepsilon\cosh\kappa_{p}(H-z)+\varepsilon_{w} p\sinh\kappa_p(H-z))(\kappa_p\varepsilon\cosh\kappa_{p}z^{\prime}+\varepsilon_{w} p\sinh\kappa_pz^{\prime})}{\varepsilon \kappa_p ((\kappa_p^2\varepsilon^2+\varepsilon_{w}^2p^2)\sinh\kappa_pH + 2\kappa_p p\varepsilon\varepsilon_{w}\cosh\kappa_pH)}, &z>z'.
\end{cases}
\end{equation}
The mean field contribution to disjoining pressure in this case is also determined by eq. (\ref{disj_press_MF}). After similar, but more cumbersome calculations, the correlation contribution to the disjoining pressure takes the following form:
\begin{equation}
\label{Pi_corr_diel}
\Pi_{cor} = -\frac{k_{B}T}{2\pi}\int\limits_{0}^{\infty}dp\,p\Delta(p;H)\frac{\kappa_p}{e^{2\kappa_p H}-1},
\end{equation}
where 
\begin{equation}
\Delta(p;H)=\frac{(\kappa_p\varepsilon-p\varepsilon_{w})^2}{\kappa_p^2\varepsilon^2+\varepsilon_{w}^2 p^2 + 2\kappa_pp\varepsilon\varepsilon_{w}\coth\kappa_pH}.
\end{equation}
It is evident that the case of metal walls is encompassed within the formula (\ref{Pi_corr_diel}). To derive the formula (\ref{Pi_corr_metal}), we need to take the formal limit $\varepsilon_{w}\to \infty$ in (\ref{Pi_corr_diel}). In the absence of ions within the slit-like pore ($\kappa=0$), eq. (\ref{Pi_corr_diel}) leads to the following classical van der Waals force per unit area between dielectric walls
\begin{equation}
\Pi_{cor}\simeq -\frac{k_{B}T\zeta(3)}{8\pi H^3}\left(\frac{\varepsilon_w-\varepsilon}{\varepsilon_w+\varepsilon}\right)^2.
\end{equation}
The latter result is in accordance with the fact that static polarizability of the dielectric walls with permittivity $\varepsilon_w$, immersed in dielectric fluid with permittivity $\varepsilon$ is proportional to $(\varepsilon_w-\varepsilon)/(\varepsilon_w+\varepsilon)$ (see, for instance,~\cite{landau2013electrodynamics,israelachvili2011intermolecular}). In the limit $\varepsilon \to \infty$, that is, when two identical dielectrics are separated by a metal layer, the compressive force acting on the layer is
\begin{equation}
\Pi_{cor}\simeq -\frac{k_{B}T\zeta(3)}{8\pi H^3}.
\end{equation}

Integral (\ref{Pi_corr_diel}) can also be evaluated in the limit $\kappa H\gg 1$. In this case the basic contribution to the integral brings $p/\kappa\ll 1$, so that we can use the approximation $\kappa_p H\simeq \kappa H +{p^2H}/{2\kappa}$ and $\Delta(p;H)\simeq \Delta(0;H)\simeq 1$. Calculating the integral, we obtain
\begin{equation}
\label{Pi_corr_diel_}
\Pi_{cor} \simeq -\frac{k_{B}T\kappa^2}{4\pi H}e^{-2\kappa H}.
\end{equation}
The same estimation for the asymptotic of the correlation contribution to the disjoining pressure was derived in ref.\cite{jancovici2004screening} using a different method.

\section{Conclusion}
Despite the milestones achieved using the thermomechanical approach in describing mechanical stresses in liquid-phase electrolytes, there are still several issues that need to be addressed in the near future when it comes to ionic fluids under nano-confinement. First, the thermomechanical approach has been applied to obtain stress tensors within the framework of mean field models of ionic fluids, which are based on 'local' thermodynamic potentials such as free energy functionals containing spatial derivatives of order parameters up to second order. However, it is still unclear how this approach can be extended to more sophisticated theories based on the free energy functionals that depend on weighted order parameters~\cite{de2020interfacial,de2022structural,de2022polar}, such as the case of hard sphere models~\cite{roth2010fundamental}. However, progress can be made by considering the case where the weighting function is the Green's function of a certain differential operator. The latter would be considered in detail elsewhere. Furthermore, it was possible to calculate the stress tensor for essentially nonlocal functionals that describe the fluctuation correction to the mean-field thermodynamic potential within statistical field theory~\cite{budkov2023variational,brandyshev2023statistical}. It is also important to apply the thermomechanical approach to ionic fluids confined in narrow pores, performing sophisticated calculations for electrostatic Green's functions~\cite{colla2016charge,dos2017simulations,girotto2017simulations} and then calculating the components of the correlation stress tensor. Second, the thermomechanical approach allows us to calculate mechanical stresses not only in ionic liquids confined in charged pores with a slit geometry (where we can use the contact theorem~\cite{dean2003field,buyukdagli2023impact,mallarino2015contact}), but also in pores of arbitrary geometry, including those with nonconvex geometries and corrugated walls~\cite{kolesnikov2022electrosorption,ruixuan2023electrostatic,nesterova2025mechanism}. Till now, only a couple of simple cases have been considered. However, the use of a thermomechanical approach for the self-consistent calculation of local stresses in complex porous materials impregnated with liquid electrolytes holds obvious potential for practical applications.

\section*{Acknowledgments}
The authors are grateful to Referee for insightful comments and valuable suggestions, which made it possible to significantly improve the original variant of the text. The authors thank the Russian Science Foundation (Grant No. 24-11-00096) for financial support.

\selectlanguage{english}
\bibliography{name}

\end{document}